\documentclass[a4paper,twoside]{article}
\newcommand{\affil}[1]{$^{\rm #1}$}
\usepackage[authoryear]{natbib}
\bibpunct{(}{)}{;}{a}{}{,}
\usepackage{graphicx}
\date{} 
\newcommand{\mbf}[1]{\mbox{\boldmath $#1$}}
\newcommand{\mbfs}[1]{\mbox{\scriptsize\boldmath $#1$}}

\newcommand{\bsf}[1]{\textbf{\textsf{#1}}}

\newcommand{\fdg}{\mbox{\ensuremath{.\!\!^\circ}}}%

\newcommand{\eqn}[1]{equation~(\ref{eqn:#1})}
\newcommand{\eqns}[1]{equations~(\ref{eqn:#1})}

\newcommand{\Ci}{\ensuremath{i}}
\renewcommand{\Re}{\rm Re}
\renewcommand{\Im}{\rm Im}

\newcommand{\trace}{{\rm tr}}

\newcommand{\Rotation}{{\bsf{R}}}
\newcommand{\Boost}{{\bsf{B}}}

\newcommand{\vRotation}[1][n]{\ensuremath{\Rotation_{\mbfs{\hat #1}}}}
\newcommand{\vBoost}[1][m]{\ensuremath{\Boost_{\mbfs{\hat #1}}}}

\newcommand{\rotat}{\ensuremath{\vRotation(\phi)}}
\newcommand{\boost}{\ensuremath{\vBoost(\beta)}}

\newcommand{\pauli}[1]{\ensuremath{ {\mbf{\sigma}}_{#1} }}

\newcommand{\psrchive}{{\sc psrchive}}
\newcommand{\psrfits}{{\sc psrfits}}

\title{\large\bf\flushleft \psrchive\ and \psrfits: 
       Definition of the Stokes Parameters and Instrumental Basis Conventions}

\author{\parbox{\textwidth}{\flushleft
\vspace{-0.5cm}
{\it W. van Straten\affil{A,C},
     R. Manchester\affil{B},
     S. Johnston\affil{B},
     and J. Reynolds\affil{B}}\\
\vspace{0.4cm}
{\small \affil{A}\,Swinburne University of Technology, PO Box 218, Hawthorn VIC 3122, Australia}\\
{\small \affil{B}\,CSIRO Astronomy and Space Science, PO Box 76, Epping NSW 1710, Australia} \\
{\small \affil{C}\,Email: vanstraten.willem@gmail.com}}}

\begin{document}
\twocolumn[
\begin{changemargin}{.8cm}{.5cm}
\begin{minipage}{.9\textwidth}
\vspace{-1cm}
\maketitle
%
%
\small{\bf Abstract:}

This paper defines the mathematical convention adopted to describe an
electromagnetic wave and its polarisation state, as implemented in the
\psrchive\ software and represented in the \psrfits\ definition.
Contrast is made between the convention that has been widely accepted
by pulsar astronomers and the IAU/IEEE definitions of the Stokes
parameters.  The former is adopted as the PSR/IEEE convention, and a
set of useful parameters are presented for describing the differences
between the PSR/IEEE standard and the conventions (either implicit or
explicit) that form part of the design of observatory instrumentation.
To aid in the empirical determination of instrumental convention
parameters, well-calibrated average polarisation profiles of
PSR J0304+1932 and PSR J0742$-$2822 are presented at radio wavelengths
of approximately 10, 20, and 40\,cm.




\medskip
\medskip
\end{minipage}
\end{changemargin}
]
\small


\section{Introduction}

Since the early days of pulsar astronomy, observations of the
polarisation properties of their radio signal have yielded a diverse range of
information about these stars and their environment, including the
properties of the pulsar emission mechanism \citep[e.g.][]{thhm71,lm88,es04},
the geometry of the pulsar magnetosphere \citep[e.g.][]{rc69a,jhv+05}, and the
properties of the Galactic magnetic field \citep[e.g.][]{man74,hml+06}.
Polarimetry also provides additional information that may be
exploited for high-precision pulsar timing \citep{van06}.

Pulsars are typically observed either at the focus of a single antenna
or at the phase centre of an antenna array.  In either case, there are
effectively two receptors that ideally respond maximally to orthogonal
senses of polarisation.  The statistical properties of the voltages
induced in the receptors are computed as a function of topocentric
pulse phase, producing a quantity known as the polarisation profile.
Well-calibrated polarisation profiles require observations of one or
more calibrator signals and a mathematical description of the
instrumental response \citep[e.g.][]{van04a}.

Even after calibration, there is often confusion regarding the
definition of left-handed and right-handed circular polarisation (LCP
and RCP) as well as the sign of Stokes~$V$ and/or the position angle
of the linear polarisation.  The uncertainty arises from the
different conventions that have persisted in the classical physics and
engineering literature.  The earliest pulsar polarisation observations
\cite[e.g.][]{man71b} established a convention that is consistent with
the Institute of Electrical and Electronics Engineers (IEEE)
definition of LCP and RCP and the definition of Stokes~$V$ by
\citet{kra66}.  This convention differs from the one later adopted by
the International Astronomical Union \citep{iau74}.  

The PSR/IEEE convention is implemented by the \psrchive\
software\footnote{http://psrchive.sourceforge.net} and encoded in the
\psrfits\ file format, both of which have been openly developed in
an effort to facilitate the exchange of pulsar astronomical data
between observatories and research groups \citep{hvm04}.  The
\psrchive\ software implements innovative methods of polarisation
calibration \citep[e.g.][]{van04a,van06} and Faraday rotation measure
estimation \citep[e.g.][]{hml+06,njkk08}.  To properly utilise these
algorithms, data must be input in a form that complies with the
PSR/IEEE convention.

In the following section, the PSR/IEEE convention is defined and
contrasted with the IAU/IEEE definitions of the Stokes parameters.
Section 3 introduces a comprehensive set of parameters that can be
used to document the differences between an instrumental design and
the PSR/IEEE convention.  In Section 4, polarisation profiles of PSR
J0304+1932 (B0301+19) and PSR J0742$-$2822 (B0740$-$28) are provided as
a point of reference for astronomers aiming to conform to the
PSR/IEEE convention.


\section{Definitions}

Consider a quasi-monochromatic electromagnetic wave with mean
frequency $\omega$, represented at the origin by the transverse
electric field vector
\begin{equation}
\mbf{e}(t) = \left( \begin{array}{c}
e_0 \\
e_1
\end{array}\right)
 = \left( \begin{array}{c}
 a_0(t) \exp i [\phi_0(t) + \omega t] \\
 a_1(t) \exp i [\phi_1(t) + \omega t]
\end{array}\right).
\label{eqn:vector}
\end{equation}
Note that the complex argument {\bf increases} linearly with time;
this sign convention is commonly encountered in engineering texts
\citep[e.g.][]{pap62,bra65,kra66} and is implicit in the definition 
of most forward discrete Fourier transform (DFT) implementations
\citep[e.g.][]{fj05}.  It is also adopted in a seminal series of
papers on radio polarimetric calibration \citep{hbs96,shb96,hb96}.
Given the above definition, time delays correspond to {\bf negative}
values of the phase $\phi$.

The polarisation of an electromagnetic wave is described by the
second-order statistics of $\mbf{e}$, as represented using the complex
$2\times2$ coherency matrix \citep{bw80}
\begin{equation}
\label{eqn:coherency}
\mbf{\rho}=\langle\mbf{e}\,\mbf{\otimes\,e}^\dagger\rangle
= \left( 
    \begin{array}{cc}
      \langle e_0e_0^*\rangle & \langle e_0e_1^*\rangle \\
      \langle e_1e_0^*\rangle & \langle e_1e_1^*\rangle
    \end{array}
  \right).
\end{equation}
Here, the angular brackets denote an ensemble average, $\mbf\otimes$
is the matrix direct product, and $\mbf{e}^\dagger$ is the Hermitian
transpose of $\mbf{e}$.  The coherency matrix may be expressed as a
linear combination of Hermitian basis matrices,
\begin{equation}
{\mbf\rho} = {1\over2}\sum_{k=0}^3 S_k\pauli{k}
 = (S_0\,\pauli{0} + \mbf{S\cdot\sigma}) / 2,
\label{eqn:rho}
\end{equation}
where $S_0$ is the total intensity, $\mbf{S} = (S_1,S_2,S_3)$ is the
Stokes polarisation vector, $\pauli{0}$ is the $2\times2$ identity
matrix, and $\mbf{\sigma} = (\pauli{1},\pauli{2},\pauli{3})$ is a
three-vector with components equal to the Pauli matrices.  The four
Hermitian basis matrices, consisting of the identity matrix and the
three Pauli matrices, are defined as
\begin{eqnarray}
\pauli{0} = \left(\begin{array}{cc}
1 & 0 \\
0 & 1 
\end{array}\right)
&
\pauli{1} = \left(\begin{array}{cc}
1 & 0 \\
0 & -1 
\end{array}\right)
\nonumber
\\
\pauli{2} = \left(\begin{array}{cc}
0 & 1 \\
1 & 0 
\end{array}\right)
&
\pauli{3} = \left(\begin{array}{cc}
0 & -i \\
i & 0
\end{array}\right).
\label{eqn:pauli}
\end{eqnarray}
The Stokes parameters are the projections of the coherency matrix
onto the basis matrices, given by
\begin{equation}
S_k = \trace(\pauli{k}\mbf{\rho}),
\label{eqn:stokes}
\end{equation}
where $\trace()$ is the matrix trace operator.

Linear transformations of the electric field are represented by
complex $2\times2$ Jones matrices.  Under the operation,
$\mbf{e}^\prime=\bsf{J}\mbf{e}$, the coherency matrix is subjected to
a congruence transformation,
${\mbf{\rho}^\prime}=\bsf{J}\mbf{\rho}\bsf{J}^\dagger$.  Using the
axis-angle parameterisation \citep{bri00}, the polar decomposition of
a Jones matrix \citep{ham00} is expressed as
\begin{equation}
\label{eqn:polar}
\bsf{J} = J \, \boost \, \rotat,
\end{equation}
where $J=|\bsf{J}|^{1/2}$, $|\bsf{J}|$ is the determinant of \bsf{J}, 
\boost\ is positive-definite Hermitian, and
\rotat\ is unitary; both \boost\ and \rotat\ are unimodular.  Under
the congruence transformation of the coherency matrix, the Hermitian
matrix
\begin{equation}
\label{eqn:Boost}
\boost = \pauli{0}\cosh\beta + \mbf{\hat{m}\cdot\sigma}\sinh\beta
\end{equation}
effects a Lorentz boost of the Stokes four-vector along the $\mbf{\hat m}$
axis by a hyperbolic angle $2\beta$.
As the Lorentz transformation of a spacetime event mixes temporal and
spatial dimensions, the polarimetric boost mixes total and polarised
intensities, thereby altering the degree of polarisation.
In contrast, the unitary matrix
\begin{equation}
\label{eqn:Rotation}
\rotat = \pauli{0}\cos\phi + \Ci\mbf{\hat{n}\cdot\sigma}\sin\phi
\end{equation}
rotates the Stokes polarisation vector about the $\mbf{\hat n}$ axis by
an angle $2\phi$.
As the orthogonal transformation of a vector in Euclidean space
preserves its length, the polarimetric rotation leaves the total
intensity and degree of polarisation unchanged.


\subsection{The IAU/IEEE Convention}

As summarised by \cite{hb96} and depicted in Figure~1, the IAU/IEEE
definitions of the Stokes parameters are based on a right-handed
Cartesian coordinate system, in which the plane wave propagates toward
the observer in the positive $z$ direction, and $e_0=e_x$ and
$e_1=e_y$ are the components of the electric field projected onto
North and East, respectively.  Noting that the sign of the complex
argument in \eqn{vector} corresponds to the upper sign convention in
equations (2) through (4) of \cite{hb96}, the Stokes parameters are
defined by
\begin{eqnarray}
\label{eqn:stokes_def}
I & = & \langle |e_x|^2 + |e_y|^2 \rangle \\
Q & = & \langle |e_x|^2 - |e_y|^2 \rangle \\
U & = & \langle 2 \Re[ e_x e_y^* ] \rangle \\
V & = & \langle 2 \Im[ e_x e_y^* ] \rangle
\end{eqnarray}
Stokes $V$ is positive for RCP, for which $e_y$ lags $e_x$ and,
looking toward the source, the observer sees the electric field rotate
counter-clockwise.  The position angle of the linearly polarised
component is also measured in a counter-clockwise direction from North
toward East.

\begin{figure*}
\label{fig:frame}
\includegraphics[angle=-90,width=76mm]{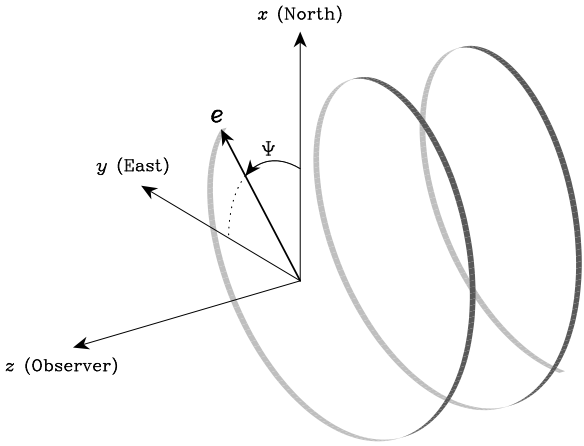}
\caption{A right-hand circularly polarised (RCP) wave in the adopted
coordinate system.  The $x$ and $y$ axes lie in the plane of the sky,
with the $x$-axis pointing North, the $y$-axis pointing East, and the
$z$-axis pointing toward the observer.
The electric field vector \mbf{e} lies in the $x$--$y$ plane; its
position angle $\Psi$ is measured with respect to North, increasing
toward East.
As the RCP wave propagates in the positive $z$ direction, the position
angle increases; that is, the electric field vector rotates
counter-clockwise, as seen by the observer.
Note that an RCP wave, as defined by the IEEE, forms a left-handed
helix in space.}
\end{figure*}


\subsection{The PSR/IEEE Convention}

In the majority of published pulsar polarisation observations,
Stokes $V$ is positive for LCP as defined by the IEEE
\citep[e.g.][]{
man71b,  
scr+84,  
xkj+98}. 
This is the convention described by \cite{kra66}; 
it is encoded in the sign of the phase in \eqn{vector}, the definition
of the Pauli matrices in \eqn{pauli}, and the relationships between
the Stokes parameters and the coherency matrix defined by \eqns{rho}
and~(\ref{eqn:stokes}).  Using either \eqn{coherency} or \eqn{rho} with
$\mbf{S}=(Q,U,V)$, it can be shown that in the linear basis,
\begin{equation}
\mbf{\rho} = {1\over2}
\label{eqn:coherency_linear}
\left( \begin{array}{cc}
      I+Q  & U-iV \\
      U+iV & I-Q
\end{array} \right).
\end{equation}
In this case,
\begin{equation}
U+iV = 2\langle e_y e_x^* \rangle.
\end{equation}
That is, Stokes $V$ is positive when the phase of $e_y$ {\bf leads}
that of $e_x$, which is opposite to the IAU convention for Stokes $V$.
The PSR/IEEE convention is adopted and used in the remainder of this
paper; it is the standard implemented by the \psrchive\ software and
expressed in the \psrfits\ definition.

The PSR/IEEE convention for the position angle of the linear
polarisation is consistent with the IAU/IEEE convention.
However, as discussed in detail by \citet{ew01}, the rotating
vector model \citep[RVM;][]{rc69a} has often been applied to
measurements using an equation in which the position angle is assumed
to increase in a {\it clockwise} direction on the sky, which is
opposite to the IAU/IEEE convention.  Therefore, care must be taken
when comparing RVM parameter estimates from different experiments.

Given the definition of the Stokes parameters in terms of linearly
polarised receptors, it is possible to also define a standard basis
for circularly polarised feeds.  The basis transformation, \bsf{C},
is chosen to effect a convenient cyclic permutation of the Stokes
parameters; that is, under the congruence transformation,
$\mbf{\rho}^\prime=\bsf{C}\mbf{\rho}\bsf{C}^\dagger$,
the Stokes vector, $\mbf{S}^\prime=(V,Q,U)$.

The transformation with unit determinant that effects
the desired cyclic permutation of the Stokes parameters is
\begin{equation}
\label{eqn:circular}
\bsf{C} = {1\over\sqrt{2i}}
\left( \begin{array}{cc}
  1 & -i \\
  1 & i 
\end{array} \right) = ( \mbf{e}_L\; \mbf{e}_R )^\dagger ,
\end{equation}
where
\begin{eqnarray}
\mbf{e}_L = {1\over\sqrt{2i}}\left( \begin{array}{c}
1 \\
i
\end{array}\right)
&
\mathrm{and}
&
\mbf{e}_R = {1\over\sqrt{2i}}\left( \begin{array}{c}
1  \\
-i 
\end{array}\right) \nonumber
\end{eqnarray}
are the Jones vectors of left and right circularly polarised receptors
expressed in the linear basis.  Apart from an absolute phase factor of
$e^{-i\pi/4}$ chosen to yield a basis transformation with unit
determinant, the above definition of circularly polarised receptors is
identical to that of \cite{hb96}.


\section{Instrumental Conventions}

A wide variety of different conventions, both explicit definitions and
implicit assumptions, are encountered in the design of observatory
instrumentation and data analysis software.  Therefore, it is
desirable to define a set of parameters that describe any known
differences between the experimental design and the chosen standard.
These corrections can be be broadly classified into four groups:
projection, basis, instrument, and backend.

The known {\bf projection} of the ideal feed receptors onto the plane
of the sky may include effects such as the parallactic rotation of an
altitude-azimuth mounted antenna, or the foreshortening of fixed
dipoles in an array.
The ideal {\bf basis} describes a feed consisting of two orthogonally
polarised receptors with unit gain.  The feed may also be equipped
with an internal reference source, such as a noise diode, that may be
used to calibrate the differential gain and phase of the
instrumentation.
The {\bf instrument} parameterises the transformation from the
receptors to the backend, as determined using a separate calibration
procedure \citep[e.g.][]{van04a}.
The {\bf backend} includes all hardware and software involved in
reducing the astronomical data to produce results that are suitable
for further analysis.
Only the {\bf basis} and {\bf backend} correction parameterisations
are discussed here.

\subsection{Basis Parameters}

In principle, the polarisations of both an ideal feed and the
reference source could be specified by four rotations; e.g. two sets
of ellipticity and orientation angles \citep{cha60}.  However, basis
rotations can be problematic due to singularities at the poles.
Furthermore, it may be preferable to use less abstract, more commonly
encountered, and/or more readily measured quantities.  The convention
may also be simplified by assuming that the receptors are either
linearly or circularly polarised, and that the reference source is
linearly polarised.  Under these assumptions, the following four
parameters are sufficient to completely describe the receiver.

\vspace{2mm}
\noindent
{\bf Feed Basis}: circularly or linearly polarised receptors

\vspace{2mm}
\noindent
{\bf Feed Hand}: left or right; if left, $e_0$ and $e_1$ are switched

\vspace{2mm}
\noindent
{\bf Symmetry Angle}: orientation of a linearly polarised signal that
  produces an equal and in-phase response in each receptor

\vspace{2mm}
\noindent
{\bf Calibrator Phase}: differential phase of the internal 
  reference source, $\Phi_c=\tan^{-1}(S_3/S_2)$

\vspace{2mm}
In the linear basis, a linearly polarised wave oscillating in the
Northeast-to-Southwest plane (positive Stokes $U$) will produce
equal and in-phase responses in each receptor; therefore, the {\it
symmetry angle} has a nominal value of 45 degrees.  Also, because the
reference source is linearly polarised ($S_3=V=0$), there are only two
possible values for the {\it calibrator phase}: 0 ($U>0$) and 180
($U<0$) degrees.

In the circular basis, a linearly polarised wave oscillating in the
North-to-South plane (positive Stokes $Q$) will produce equal and
in-phase responses in each receptor; therefore, the {\it symmetry
angle} has a nominal value of 0 degrees.  Also, because $S_2=Q$ and
$S_3=U$, the {\it calibrator phase} is arbitrary.

Given these parameters, the Jones matrix of the feed is given by
\begin{equation}
\bsf{J} = \bsf{X} \bsf{C} \vRotation[v](\Theta),
\end{equation}
where \bsf{X} is the identity in a right-handed system and the
exchange matrix in a left-handed system, \bsf{C} is the identity in
the linear basis and given by \eqn{circular} in the circular basis,
and $\Theta$ is equal to the {\it symmetry angle} less its nominal
value (45 degrees in the linear basis and 0 degrees in the circular
basis).  Note that, as defined by \eqn{Rotation},
$\vRotation[v](\Theta)$ effects a rotation of the Stokes polarisation
vector about the Stokes $V$ axis by an angle $2\Theta$.
The coherency matrix of the internal reference source is 
\begin{equation}
\mbf{\rho_c} = {1\over2}\left( \begin{array}{cc}
1 & \exp (-i\Phi_c) \\
\exp (i\Phi_c) & 1 
\end{array}\right),
\end{equation}
where $\Phi_c$ is the calibrator phase.

\subsection{Backend parameters}

To describe the design of the backend, it is sufficient to consider
only the complex conjugation of the electric field.  This occurs
during lower sideband down conversion and/or when the design of an
instrument is based upon a different convention for the sign of the
phase than that adopted in \eqn{vector}.  In the case of dual-sideband
down conversion, complex conjugation occurs when the phase of the
quadrature component leads that of the in-phase component of the
signal.  Complex conjugation causes a sign change in $S_3$, a
reflection that cannot be represented by a Jones matrix.  The
following two parameters are used to describe the backend.

\vspace{2mm}
\noindent
{\bf Backend Phase}: positive or negative

\vspace{2mm}
\noindent
{\bf Down Conversion}: true or false; if true, the conjugation
  due to lower sideband down conversion has been corrected

\vspace{2mm}
\noindent
The correction of complex conjugation is performed when either 
of the following conditions is exclusively true:
\begin{enumerate}
\item[A)] {\it backend phase} is negative, or 
\item[B)] {\it down conversion} is false and lower sideband down 
	conversion was used.
\end{enumerate}
The correction is not performed when both A
and B are true because they negate each other.

\subsection{Implementation}

Each of the parameters described in the preceding sections has a
corresponding parameter in the \psrfits\ definition and a representation
in the \psrchive\ software.  Table~\ref{tab:parameters} summarises these
parameters, the range of acceptable values for each parameter, and the
effects that they have on calibrated data.

\begin{table*}
\begin{center}
\caption{Instrumental convention parameters and their effects in each basis.}
\label{tab:parameters}
\begin{tabular}{lccc|c}
\hline

Parameter           & \psrfits  & Range        & \multicolumn{2}{c}{Effect} \\
                    & Name      &              & Linear & Circular          \\

\hline

Backend Phase       & BE\_PHASE & $\pm$1       & $\pm$ V & $\pm$ U       \\

Down conversion$^a$ & BE\_DCC   & 0/1          & $\pm$ V & $\pm$ U       \\

Feed Basis          & FD\_POLN  & LIN or CIRC  & (Q,U,V) & (V,Q,U)   \\

Feed Hand           & FD\_HAND  & $\pm$ 1      & $\pm$ Q\&V & $\pm$ U\&V \\

Symmetry Angle      & FD\_SANG  & $-\pi/2<\theta<\pi/2$
& $\vRotation[v](\theta-\pi/4)$ & $\vRotation[v](\theta)$ \\

Calibrator Phase$^b$ & FD\_XYPH & $-\pi<\Phi_c<\pi$ 
& $\pm$ U\&V & $\vRotation[v](\Phi_c/2)$ \\

\hline
\end{tabular}
\medskip\\
$^a$ Applies only when down conversion is effectively lower sideband. \\
$^b$ In the linear basis, $\Phi_c = 0$ or $\pi$.
\end{center}
\end{table*}


\section{Discussion}

In general, it is non-trivial to determine the instrumental convention
parameters from first principles and an empirical approach must be
adopted.  This typically requires observations of either artificial
sources with known polarisation or astronomical sources that have been
previously well calibrated, such as the average polarisation profiles
of PSR J0304+1932 and PSR J0742$-$2822 presented in Figure~2.  These
sources are visible to observatories in both Northern and Southern
hemispheres; they also have relatively high flux densities,
well-defined position angle variation, and sufficient circular
polarisation for the purposes of comparison with other observations.

These data were recorded at the Parkes Observatory on 2009 October 9
and 10 using the 64\,m antenna and two of the Parkes Digital Filter
Banks (PDFB3 and PDFB4).  For each pulsar, a 256\,MHz band centred at
1369\,MHz was observed using the centre beam of the 20\,cm Multibeam
receiver and two bands, one with 64\,MHz of bandwidth centred at
732\,MHz and one with 1024\,MHz of bandwidth centred at 3.1\,GHz, were
observed using the 10-50\,cm dual-band receiver.  Prior to each pulsar
observation, a noise diode coupled to the receptors was observed and
used to calibrate the differential gain and phase of the instrument.
For the 20\,cm observations, the receptor cross-coupling parameters
were calibrated as described in \citet{van04a}.  Absolute flux
calibration was performed using observations of the noise diode
recorded while pointing on and off the bright Fanaroff-Riley type I
radio galaxy, 3C 218 (Hydra~A), which was assumed to have a flux
density of 43.1\,Jy at 1400\,MHz and a spectral index of 0.91.

\begin{figure*}
\label{fig:profiles}
\begin{tabular}{ll}
\includegraphics[angle=0,width=60mm]{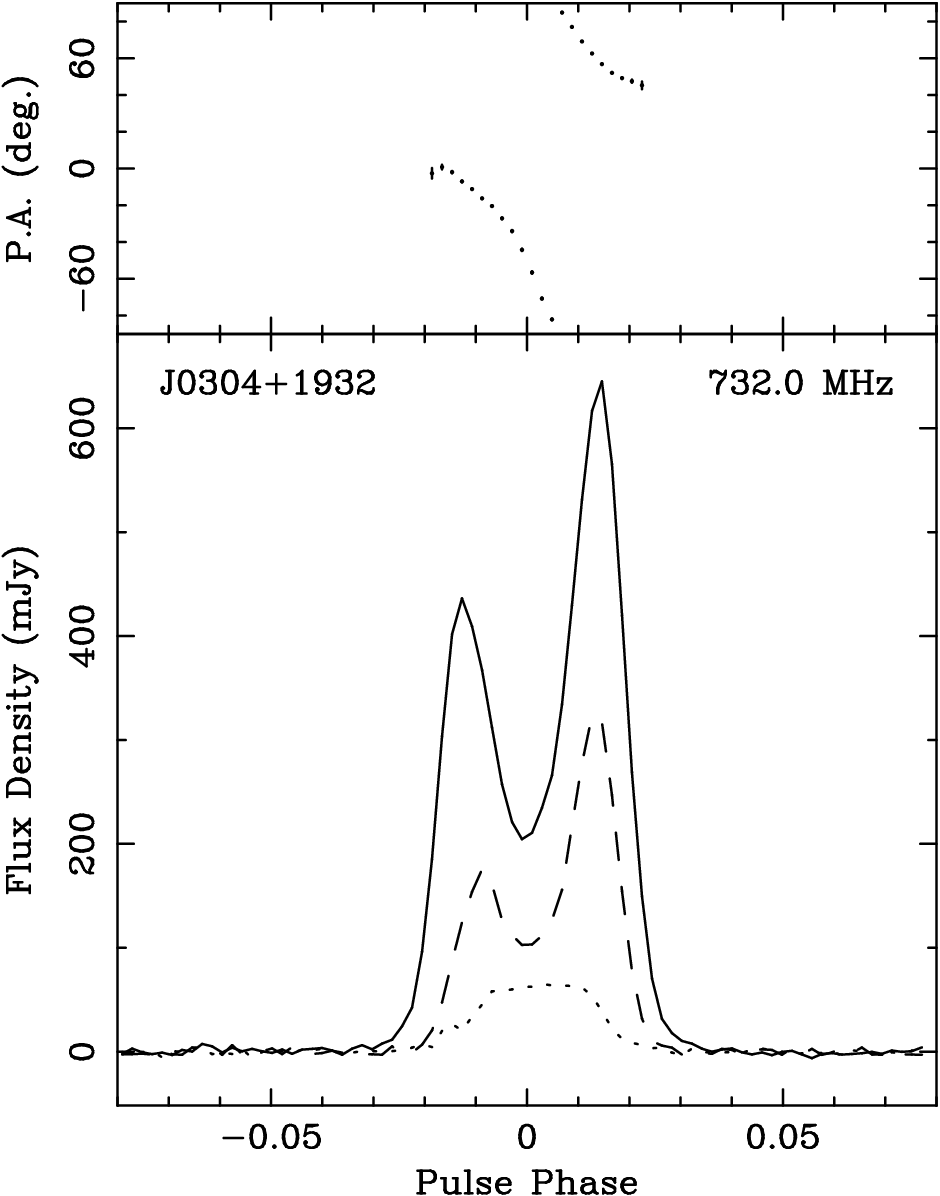} &
\includegraphics[angle=0,width=60mm]{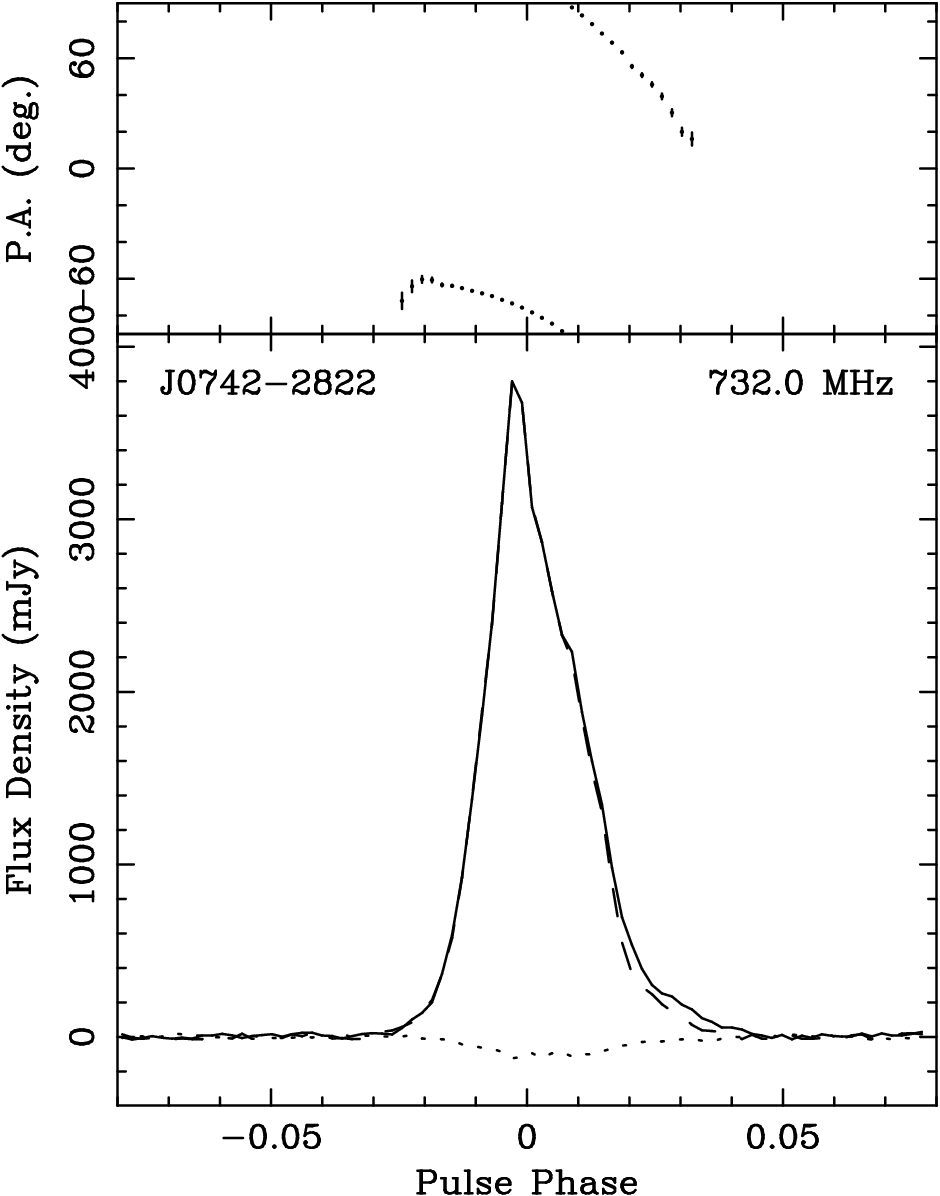} \\
\includegraphics[angle=0,width=60mm]{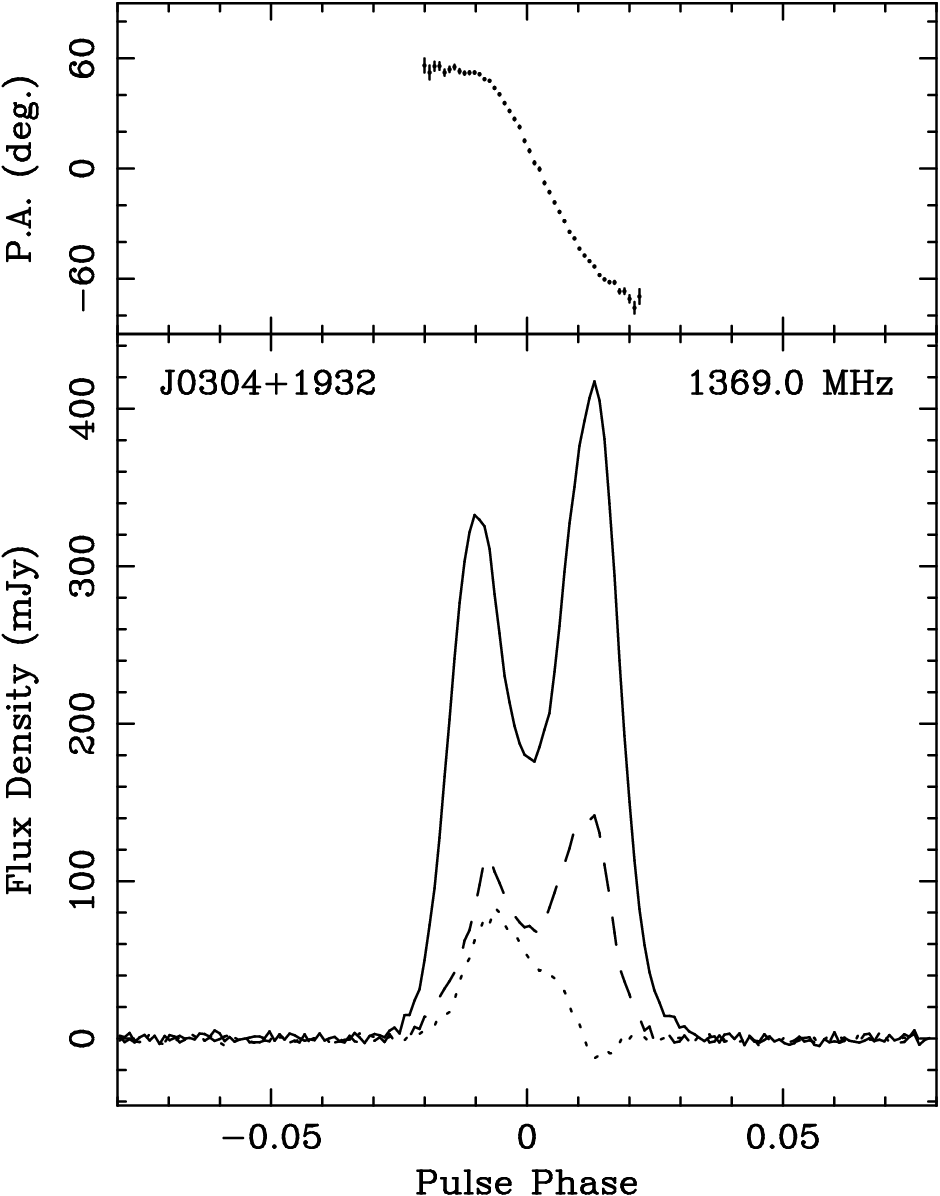} &
\includegraphics[angle=0,width=60mm]{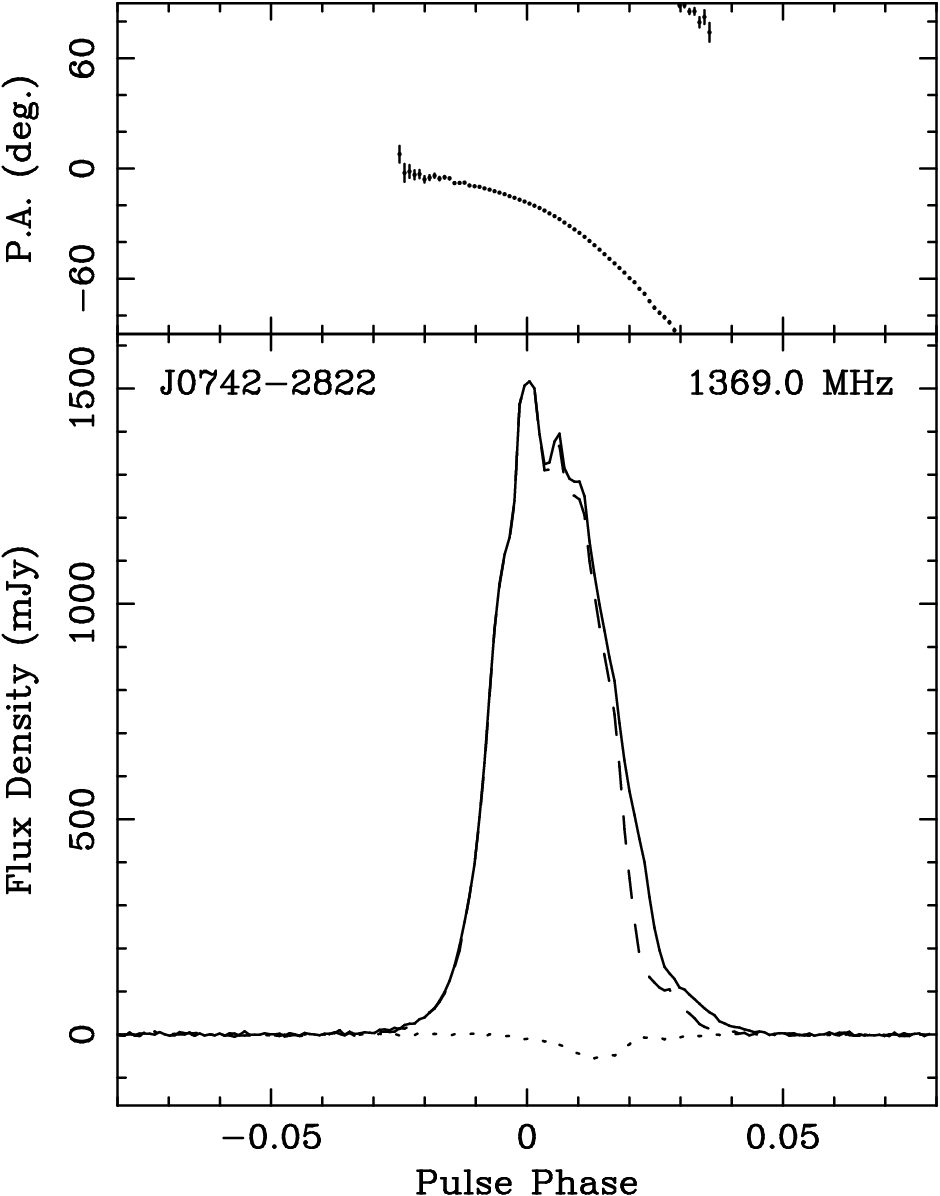} \\
\includegraphics[angle=0,width=60mm]{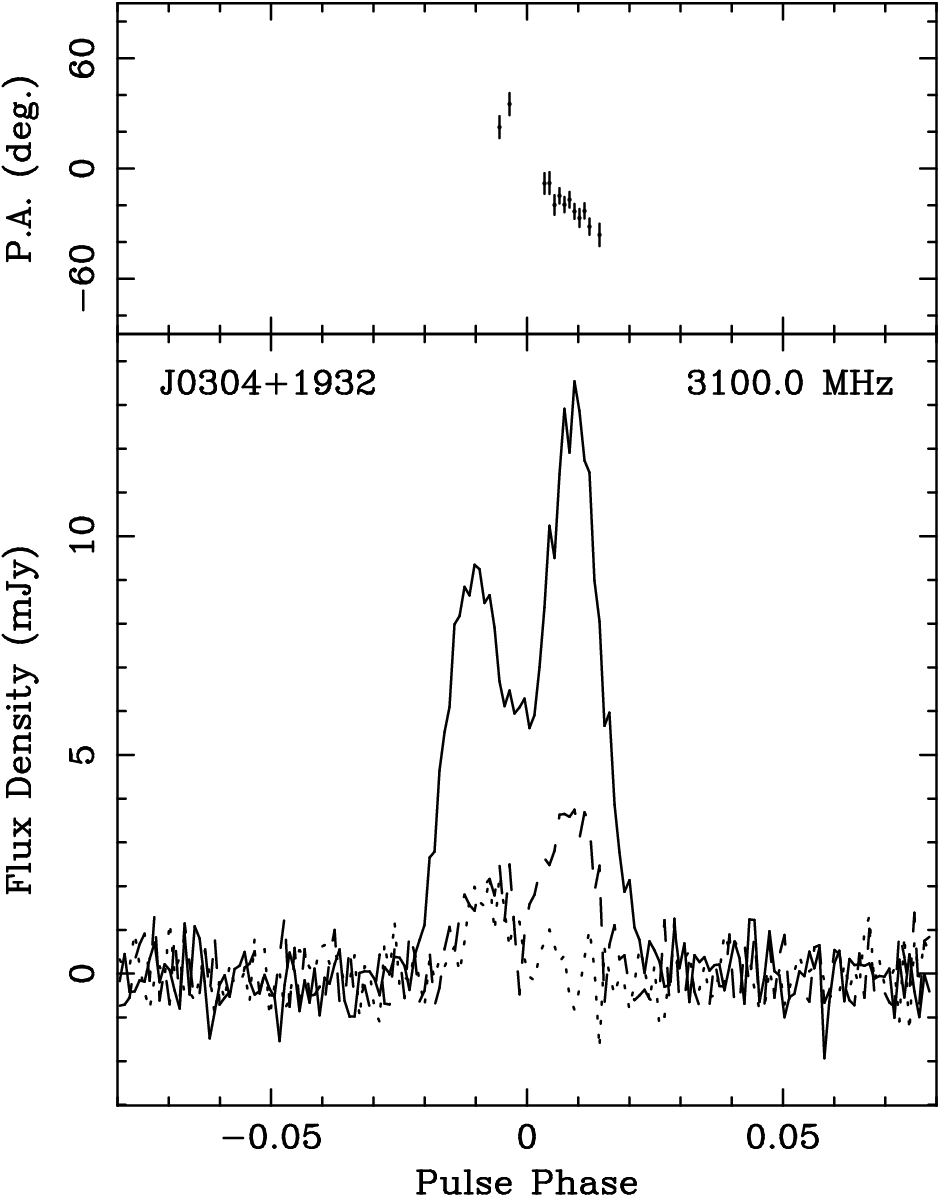} &
\includegraphics[angle=0,width=60mm]{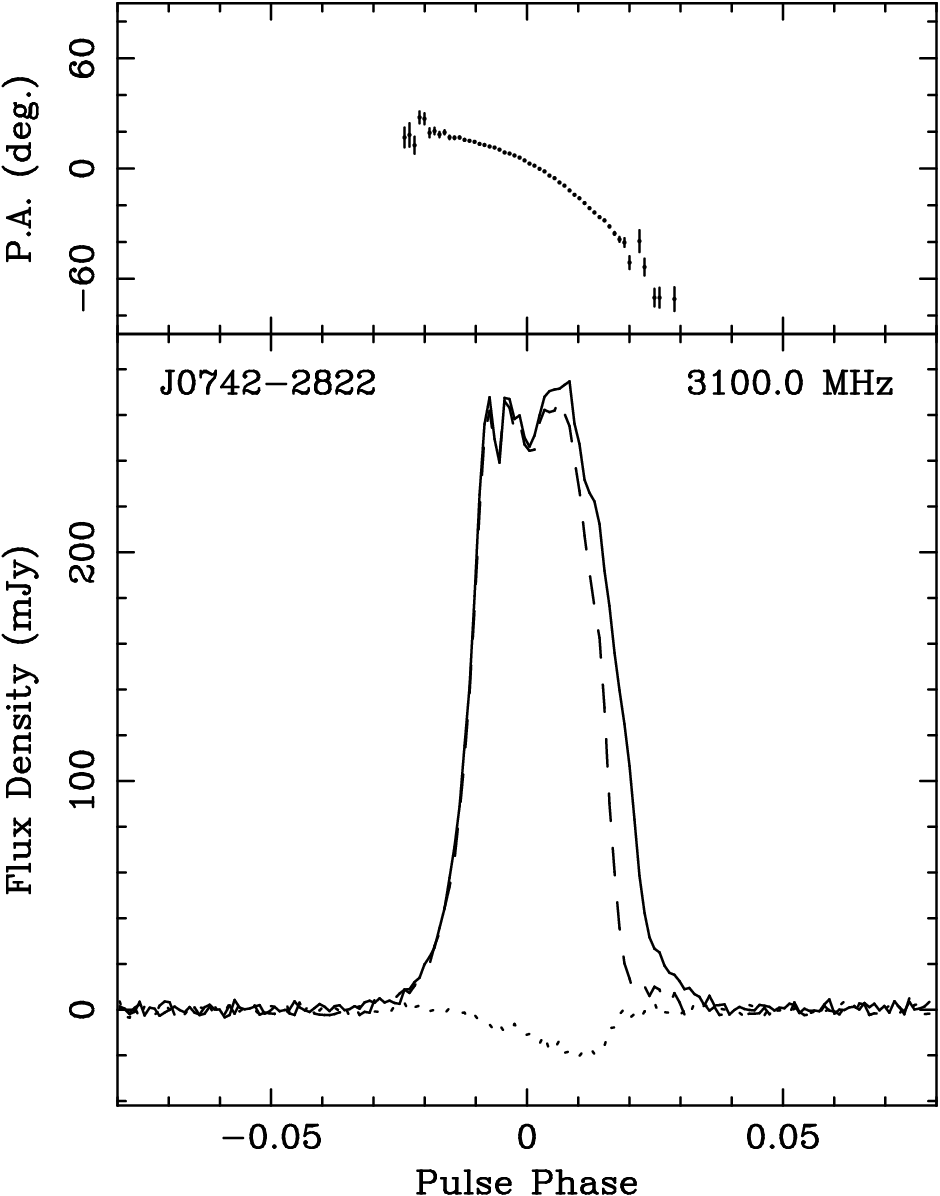}
\end{tabular}
\caption{Average polarisation profiles of PSR J0304+1932 (left) and 
PSR J0742$-$2822 (right), observed at radio wavelengths of
approximately 40\,cm (top), 20\,cm (middle), and 10\,cm (bottom).  In
each plot, the lower panel displays the total intensity, Stokes $I$
(solid); the linearly polarised flux, $L=\sqrt{Q^2+U^2}$ (dashed); and
the circular polarisation, Stokes $V$ (dotted); the upper panel plots
the position angle of the linear polarisation, $\Psi= 0.5
\tan^{-1}(U/Q)$ (the error bars are two standard deviations in
length).  Stokes~$V$ is positive for left-hand circularly polarised
radiation as defined by the IEEE.}
\end{figure*}

The data were corrected for Faraday rotation using the most recently
published rotation measure (RM) for each pulsar, as determined using
the ATNF Pulsar Catalogue \citep{mhth05}. For PSR J0304+1932,
RM=$-8.3$\,rad~m$^{-2}$ \citep{man74}; for PSR J0742$-$2822,
RM=+149.95\,rad~m$^{-2}$ \citep{jhv+05}.  The position angles plotted
in Figure~2 are referred to the centre frequency of the band and are
as observed at Parkes.  Variations in the Faraday rotation along the
line of sight, particularly those occurring in the Earth's
ionosphere, will rotate the observed angles by a small amount. The
ionospheric RM component was estimated using the International
Reference Ionosphere models \citep{bil01} and was about
$-0.7$~rad~m$^{-2}$ in the direction of PSR J0304+1932 and
$-1.5$~rad~m$^{-2}$ in the direction of PSR J0742$-$2822. Position
angles above the ionosphere were therefore greater than the plotted
values by about $0\fdg4$ and $0\fdg8$ at 3100\,MHz, $0\fdg7$ and
$1\fdg5$ at 1369\,MHz, and $7\fdg5$ and $14\fdg4$ at 732\,MHz for PSR
J0304+1932 and PSR J0742$-$2822 respectively.

Referring to the effects defined in Table~1, the instrumental
convention parameters of new systems can be adjusted until the
observed values of Stokes $V$ and position angle match those plotted
in Figure~2.  This comparison should be performed only after the {\em
projection} (e.g. parallactic rotation) and {\em instrument}
(e.g. differential gain and phase) have been calibrated.  Given the
overlapping domains of the parameter effects, the proper selection of
parameter values will most likely require consultation with both
engineering documentation and observatory staff.

\section*{Acknowledgments} 

The authors are grateful to Jonathon Kocz for observing assistance and
to Paul Demorest, Andrew Lyne, and Ingrid Stairs for helpful
discussions on this topic.  The Parkes Observatory is part of the
Australia Telescope which is funded by the Commonwealth of Australia
for operation as a National Facility managed by CSIRO.

\bibliographystyle{apj}


\end{document}